\documentclass[aps,prb,twocolumn,showpacs,amsmath,amssymb,superscriptaddress, citeautoscript]{revtex4-2}
\setcitestyle{super}
\usepackage[cmtip,all]{xy}
\usepackage{graphicx}
\usepackage{hyperref}
\usepackage{xcolor}
\usepackage{color}
\usepackage[all,cmtip]{xy}
\usepackage{multirow}
\usepackage{natbib}
\usepackage{amsmath,amsthm} 
\usepackage{amsfonts}

\makeatother

\begin{document}

\title{Model Hamiltonian for Altermagnetic Topological Insulators}

\author{Rafael Gonz\'{a}lez-Hern\'{a}ndez}
\email{rhernandezj@uninorte.edu.co}
\affiliation{Departamento de F\'{i}sica y Geociencias, Universidad del Norte, Km. 5 V\'{i}a Antigua Puerto Colombia, Barranquilla 081007, Colombia}
\author{Bernardo Uribe}
\email{bjongbloed@uninorte.edu.co}
\affiliation{Departamento de Matem\'{a}ticas y Estad\'{i}stica, Universidad del Norte, Km. 5 V\'{i}a Antigua Puerto Colombia, Barranquilla 081007, Colombia}

\date{\today}

\begin{abstract}
	We present models of topological insulating Hamiltonians exhibiting intrinsic altermagnetic features, 
	protected by combined three-fold or four-fold rotational symmetries with time-reversal. 
	We demonstrate that the spin Chern number serves as a robust topological invariant in two-dimensional systems, while for three-dimensional structures, 
	the topological nature is characterized by the spin Chern numbers computed on the $k_z$=$0$ and $k_z$=$\pi$ planes.
	The resulting phases support symmetry-protected boundary modes, including corner, hinges and surface states, whose structure is determined by the magnetic symmetry and the local  magnetic moments. 
	Our findings bridge the fields of altermagnetism and topological quantum matter, and establish a theoretical framework for engineering spintronic topological systems without net magnetization.
\end{abstract}
\maketitle

\section*{Introduction}

Altermagnetism has recently emerged as a distinct class of collinear magnetic order, complementing the conventional paradigms of ferromagnetism and antiferromagnetism. 
Unlike ferromagnets, which exhibit a net magnetization, or conventional antiferromagnets, where the electronic structure remains spin-degenerate throughout the entire Brillouin zone, altermagnets combine zero net magnetic moment with a momentum-dependent spin splitting of the electronic bands \cite{Smejkal2020,Smejkal2022,Smejkal2022_2,fesb2-am,smejkal_nat_review_22,bai2024}. 
Such unconventional behavior originates from a symmetry-enforced breaking of Kramers degeneracy. 
This is enabled by a 
 time-reversal symmetry breaking, together with the presence of  a combination of specific crystalline symmetries, such as rotations or  rotoinversions, with time reversal symmetry, thus connecting magnetic sublattices with opposite spin orientation.
This momentum-space spin splitting occurs without spin-orbit coupling and it is symmetry-protected, distinguishing altermagnets from both traditional magnetic systems and spin–orbit band splitting  \cite{jiang2024,SSG-enumeration,SSG-full,Smejkal2025}. 
Recent experimental studies have provided a clear evidence for such behavior in candidate materials, as MnTe\cite{Betancourt2023,krempasky_jungwirth_nature_24,altermagnetic-MnTe,lee2024mnte}, CrSb\cite{altermagnetic-CrSb,lu2024CrSb,CrSb-ss,CrSb-topology} and Mn$_5$Si$_3$ \cite{Reichlova2024,mn5si3-PhysRevB.109.224430,mn5si3-PhysRevB.110.L220411,mn5si3-ane}, confirming the theoretical predictions and establishing altermagnetism as a viable platform for spintronic applications. 

On the other hand, the study of topological phases of matter has transformed our understanding of quantum materials  \cite{hasan2010colloquium,qi2010topological,RevModPhys.88.035005,librotopological2021}. 
Topological insulators (TIs), characterized by insulating bulk bands and robust gapless boundary states, are classified by quantized topological invariants such as the $\mathbb{Z}_2$ index or the spin Chern number \cite{kane2005z2,sheng2006quantum,fu2006time,fu2007topological,Prodan-SCN,spin-resolved-3d,Spin_Weyl_topological_insulators,PhysRevResearch.5.033013,Spin_Chern_Number_Altermagnets}. 
These phases also support dissipationless edge or surface transport channels.  
The intersection of altermagnetism and topology represents a rapidly developing area of research. 
Altermagnets, with their intrinsic spin-split bands and time-reversal symmetry breaking, offer a natural foundation for realizing topological phases with unconventional spin textures and protected edge states \cite{toposuperAM,majoranaAM,ma2024altermagnetic,floatingBHZ-AM,weakTI-AM,heinsdorf2025altermagnetic,kanemele-AM,topoAMmagnons}. 
In particular, symmetry operations of the form $\mathcal{C}_l\mathbb{T}$, where $\mathcal{C}_l$ is an $l$-fold rotation and $\mathbb{T}$ denotes time-reversal symmetry, can stabilize topological band structures with altermagnetic character. 
While previous models have focused on metallic systems featuring nodal lines or Weyl points protected by magnetic symmetries \cite{nodalines-am,ma2024quantum,parshukov2025topological,AM-in-lieb-metal,nodalineAM-z3,topo_C4T_2024,berryC4T,menghan2025diracAM}, the realization of fully gapped 3D topological altermagnets remains an open challenge. 
Engineering tight-binding models or materials that combine the $\mathcal{C}_l\mathbb{T}$ magnetic symmetries with a bulk band gap is thus a key goal for advancing both theoretical understanding and experimental discovery.

In this work, we present a framework for engineering two-dimensional (2D) and three-dimensional (3D) topological insulator Hamiltonians that exhibit altermagnetic behavior. 
The approach begins by designing a spin-up sector Hamiltonian with a non-zero Chern number and $(\mathcal{C}_l)^2$ symmetry. A full 4$\times$4 Hamiltonian is then constructed adjoining a spin-down sector preserving the full $\mathcal{C}_l\mathbb{T}$ symmetry, resulting in spin-resolved bands with distinct energies while ensuring an odd spin Chern number to retain the topological insulating phase. The 3D Hamiltonian is constructed by adding an off-diagonal term which keeps the Hamiltonian gapped and intertwines the a trivial 2D Hamiltonian with a 2D TI Hamiltonian in
differents $k_z$-planes. 

We specifically analyze two symmetry cases: $\mathcal{C}_{4z}\mathbb{T}$ and $\mathcal{C}_{3z}\mathbb{T}$, corresponding to $d$-wave and $f$-wave altermagnetic states respectively. 
For each case, we provide model Hamiltonians, explore their electronic band structures, spin textures in momentum space, edge-state spectra in real space, and topological phase diagrams using the spin Chern number as the defining invariant. 
These constructions not only elucidate the symmetry principles governing topological altermagnetic phases but also serve as a signal for material realizations in cases such as FeSe, where the interplay of symmetry, topology and magnetism could lead to discover novel quantum phases.

\section{Engineering TI altermagnetic Hamiltonians}

\subsection{2D Hamiltonians}

There is a procedure to engineer an  4$\times $4 TI Hamiltonian with altermagnetic features.
The altermagnetic structure can be obtained whenever an $l$-fold rotation $\mathcal{C}_{l}$ composed with time reversal $\mathbb{T}$ is   
enforced on a Hamiltonian and its eigenstates are not degenerate. The topological insulating
feature is enforced if the spin Chern number (SCN) is odd.

We start with a 2$\times $2 gapped Hamiltonian whose Chern number is $\pm 1$ and which
preserves the symmetry $(\mathcal{C}_l \mathbb{T})^2=-(\mathcal{C}_l)^2$. This Hamiltonian will be modelling the spin up eigenstates
and will be denoted as follows:
\begin{align} \label{definition H up}
H^{\uparrow}({\bf{k}}) =
\begin{pmatrix}
M^{\uparrow}({\bf{k}}) & A^{\uparrow}({\bf{k}})  \\
A^{\uparrow}({\bf{k}})^* & -M^{\uparrow}({\bf{k}}) 
\end{pmatrix}.
 \end{align} 
The Hamiltonian preserves the $(\mathcal{C}_l \mathbb{T})^2$ symmetry if 
\begin{align} \label{(Cl)2 symmetry}
(\mathcal{C}_l)^2 H^{\uparrow}({\bf{k}}) ( \mathcal{C}_l)^{-2} = H^{\uparrow}((\mathcal{C}_l)^2 {\bf{k}}) 
\end{align}
where $\mathcal{C}_l$ is the diagonal matrix $\mathcal{C}_l = \mathrm{diag}(e^{\frac{{2 \pi i }}{l}}, e^{-\frac{{2 \pi i }}{l}})$.
The topological insulating feature of $H^{\uparrow}$ can be secured if we choose a variant
of the BHZ Hamiltonian  for each value of $l  \in \{3,4\}$.

Define the $ 4 \times 4$ Hamiltonian
 \begin{align} \label{2D Hamiltonian}
 H({\bf{k}}) = \begin{pmatrix}
 H^{\uparrow}({\bf{k}})  & \\
 & H^{\downarrow}({\bf{k}}) 
 \end{pmatrix}
 \end{align}
where $H^{\downarrow}({\bf{k}})$  is defined by the equation
\begin{align} \label{definition H(ClT)}
H^{\downarrow}({\bf{k}}) = \mathcal{C}_l \left( H^{\uparrow}(\mathcal{C}_l \mathbb{T}{\bf{k}}) \right)^*\mathcal{C}_l^{-1}
\end{align}
and assign to the operators   $\mathbb{T}$ and $\mathcal{C}_l$ the $4 \times 4$ matrices
\begin{align} \label{T_and_C_l}
\mathbb{T} \mapsto   \begin{pmatrix}
0 & \mathrm{I}_{2} \\
- \mathrm{I}_2 &  0  
 \end{pmatrix} \mathbb{K}
 \ \ \mathrm{and} \ \ \ 
 \mathcal{C}_l \mapsto  \begin{pmatrix}
\mathcal{C}_l & 0 \\
0&  \mathcal{C}_l^{-1}  
 \end{pmatrix} 
\end{align}
respectively. The Hamiltonian $H({\bf{k}})$ is built to preserve the symmetry $\mathcal{C}_l \mathbb{T}$, which in equations  is the following,
\begin{align}
(\mathcal{C}_l \mathbb{T}) H ({\bf{k}}) (\mathcal{C}_l\mathbb{T})^{-1} =H(\mathcal{C}_l \mathbb{T} {\bf{k}}),
\end{align}
and this equation is satisified if and only if Eqns.   \eqref{(Cl)2 symmetry} and \eqref{definition H(ClT)} 
are satisfied. Therefore we just need to check Eqn. \eqref{(Cl)2 symmetry}  for the 2$\times$2
Hamiltonian $H^\uparrow$ and Eqn.   \eqref{definition H(ClT)} specifies how to construct the Hamiltonian $H^\downarrow$.

Here the spin $z$ matrix $S_z$ is the diagonal matrix $S_z= \mathrm{diag}(1,1-1-1)$ and
it commutes with the Hamiltonian $H ({\bf{k}})$.

If $E_0^{\uparrow}({\bf{k}})$ and $E_0^{\downarrow}({\bf{k}})$ are the negative energies
of the Hamiltonians $H^{\uparrow}$ and $H^{\downarrow}$ respectively, we know from Eqn.
\eqref{definition H(ClT)}  that $E_0^{\uparrow}({\bf{k}}) = E_0^{\downarrow}(\mathcal{C}_l \mathbb{T}{\bf{k}})$.
The altermagnetic feature of the Hamiltonian $H$ emerges when $H^{\uparrow}$
is chosen such that $E_0^{\uparrow}({\bf{k}}) \neq E_0^{\downarrow}({\bf{k}})$.
Whenever this happens, the spin up valence eigenstate $\psi_0^{\uparrow}({\bf{k}})$ 
has a different energy than the one of the spin down valence eigenstate $\psi_0^{\downarrow}({\bf{k}})$,
and the altermagnetic feature of the Hamiltonian appears.

\subsubsection{$\mathcal{C}_{4z} \mathbb{T}$ symmetry}

The case $l$=$4$ has been studied in several works with appealingly similar Hamiltonians \cite{C4T-10.21468/SciPostPhys.15.3.114, ma2024altermagnetic, Spin_Chern_Number_Altermagnets}.
Here we propose a very general realization of a TI Altermagnetic Hamiltonian with $\mathcal{C}_{4z}\mathbb{T}$ symmetry which incorporates the parameters of Hamiltonians appearing in previous works \cite{C4T-10.21468/SciPostPhys.15.3.114, ma2024altermagnetic, Spin_Chern_Number_Altermagnets}.

Consider the Hamiltonian $H_4^\uparrow({\bf{k}})$ of BHZ \cite{BHZ} type   with:
 \begin{align} 
 M_4^{\uparrow}({\bf{k}}) & = M_0 - K_1 \cos (k_x) - K_2\cos(k_y),\\
 A_4^{\uparrow}({\bf{k}})  & = G_1 \sin(k_x) + i G_2 \sin (k_y),
 \end{align}
where $K_1,K_2,G_1,G_2$ are non-zero real parameters. 
Whenever $K_i$=$1$=$G_i$ we obtain the BHZ  Hamiltonian which is gapless for $M_0 \in \{-2,0,2\}$
and whose Chern number is
\begin{align}
c_1 = \left\{ \begin{matrix} \label{Chern_number_C4}
0 & \mathrm{for} &  |M_0| >2 \\
-1 & \mathrm{for} & 0 < M_0 <2 \\
1 & \mathrm{for}  & -2 < M_0 <0.
\end{matrix} \right.
\end{align}

The $(\mathcal{C}_{4z})^2=\mathcal{C}_{2z}$ action on momentum coordinates is $\mathcal{C}_{2z}(k_x,k_y)$=$(-k_x,-k_y)$ and the 2$\times $2
associated matrix is $ \mathrm{diag}(i,-i)$. Eqn. \eqref{(Cl)2 symmetry} is satisfied because of the following identities:
\begin{align}
(\mathcal{C}_{4z})^2 H_4^{\uparrow}({\bf{k}}) ( \mathcal{C}_{4z})^{-2}  = & \mathrm{diag}(i,-i) H_4^{\uparrow}({\bf{k}}) \mathrm{diag}(-i,i)\\
=& H_4^{\uparrow}(-{\bf{k}}) \\
= & H_4^{\uparrow}((\mathcal{C}_{4z})^2{\bf{k}}).
\end{align}

Since $\mathcal{C}_{4z} \mathbb{T}(k_x,k_y)=(k_y,-k_x)$, Eqn. \eqref{definition H(ClT)} reads
\begin{align}
H_4^{\downarrow}(k_x,k_y) =&\\
 \mathrm{diag}(\tfrac{1+i}{\sqrt{2}},\tfrac{1-i}{\sqrt{2}})& \left( H_4^{\uparrow}(k_y,-k_x) \right)^*  \mathrm{diag}(\tfrac{1-i}{\sqrt{2}},\tfrac{1+´i}{\sqrt{2}}) 
 \end{align}
which implies that
\begin{align}
 M_4^{\downarrow}({\bf{k}}) & = M_0 - K_1 \cos (k_y) - K_2\cos(k_x),\\
 A_4^{\downarrow}({\bf{k}})  & = i  A_4^{\downarrow}(k_y,-k_x)^*\\
 &  =-G_2 \sin(k_x) + i G_1 \sin (k_y).
 \end{align}

The 4$\times$4 Hamiltonian
\begin{align}\label{C4THamiltonian}
H_4({\bf{k}})= \begin{pmatrix}
H_4^{\uparrow}({\bf{k}})  & \\ 
& H_4^{\downarrow}({\bf{k}}) 
\end{pmatrix}
\end{align}
preserves inversion symmetry $P$, which is equivalent to $\mathcal{C}_{2z}$ in two-dimensions, breaks
times reversal symmetry and preserves the combination $\mathcal{C}_{4z} \mathbb{T}$ with associated matrix
the product 
\begin{align}\mathcal{C}_{4z} \mathbb{T} \mapsto
\begin{pmatrix}
\mathcal{C}_{4z} & \\
 & \mathcal{C}_{4z}^{-1} 
\end{pmatrix} 
\begin{pmatrix}
0 & \mathrm{I}_{2} \\
- \mathrm{I}_2 &  0  
 \end{pmatrix} \mathbb{K}
\end{align} 
where $ \mathrm{diag}(\tfrac{1+i}{\sqrt{2}},\tfrac{1-i}{\sqrt{2}})$ is the matrix associated to $\mathcal{C}_{4z}$.

\begin{figure}
	\includegraphics[width=8.5cm]{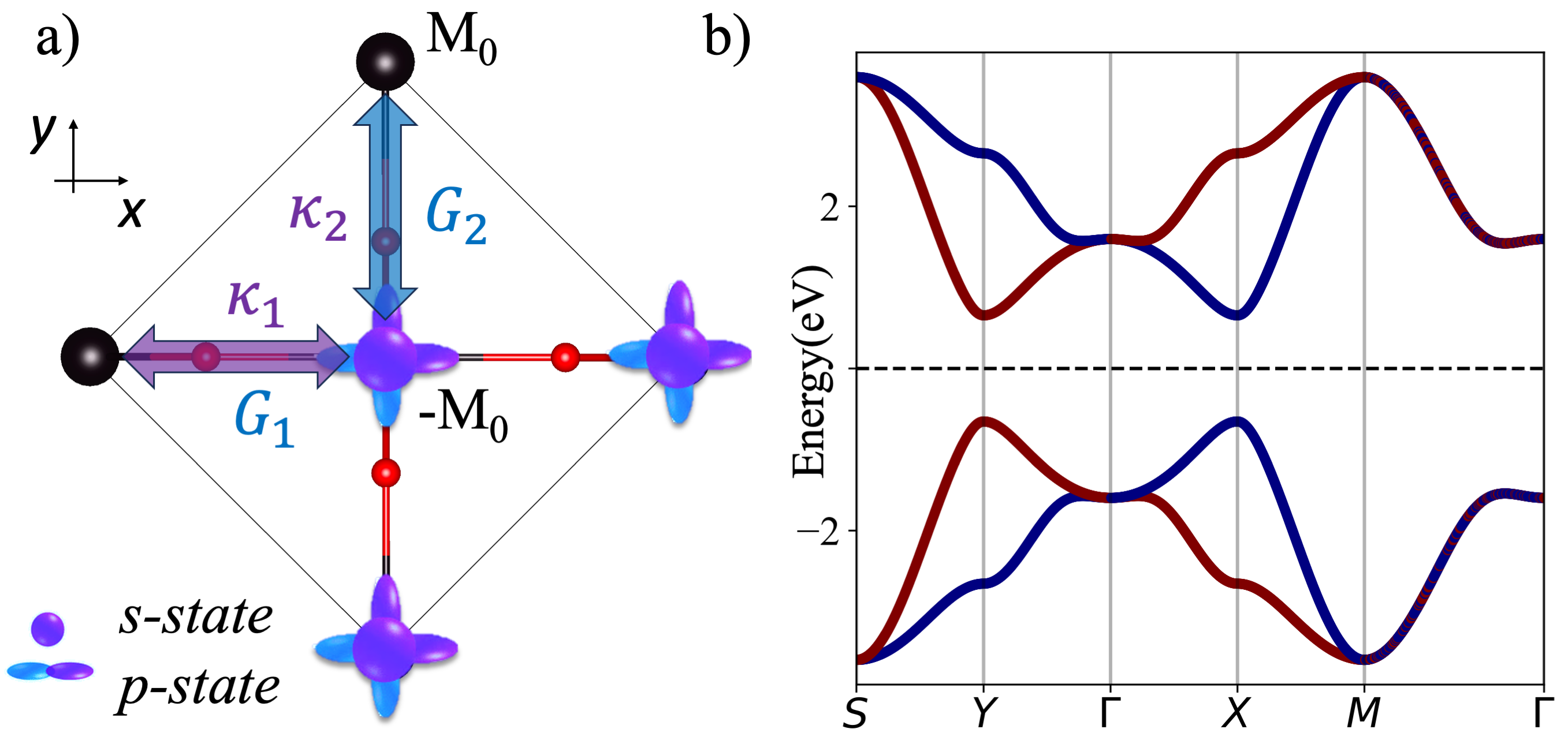}
	\caption{(a) Schematic representation of a two-dimensional $\mathcal{C}_{4z}\mathbb{T}$-symmetric altermagnetic system on an square lattice. The diagram illustrates the tight-binding model and hopping parameters, where arrows indicate the hopping interactions. The mass term $M_0$ captures the sublattice asymmetry arising from oppositely aligned local magnetizations. $K_1$ and $K_2$ denote the intra-site hoppings between orbitals ($s$–$s$ and $p$–$p$) along the $x$- and $y$-directions, respectively. $G_1$ and $G_2$ represent the inter-site hoppings (between $s$ and $p$ orbitals) along $x$ and $y$, respectively, as defined in the Hamiltonian in Eqn. \eqref{C4THamiltonian}. (b) Spin-resolved band structure for the altermagnetic model, with parameters set to $M_0=1$, $K_1$= $2.125$, $K_2$=$\tfrac{1}{K_1}$, $G_1 = 1.75$, and $G_2$=$\tfrac{1}{G_1}$. All parameters are given in $eV$.
	Time-reversal symmetry breaking leads to opposite spin splitting of the bands along the $\Gamma$–X and $\Gamma$–Y paths.}
	\label{model}
\end{figure}

Fig. \ref{model} a) illustrates the tight-binding model of a 2D altermagnetic phase on a square lattice, characterized by $\mathcal{C}_{4z}\mathbb{T}$ symmetry. 
The term $M_0$ serves as a mass term, capturing the asymmetry between the two sublattice atoms due to opposite local magnetizations that cancel out in total. 
The model includes the hopping parameters: $K_1$ and $K_2$ denote intra-orbital ($s$–$s$ and $p$–$p$) up-channel hoppings along the $x$ and $y$ directions, respectively, while $G_1$ and $G_2$ correspond to inter-orbital ($s$–$p$) up-channel hoppings along the same directions. 
For the down-channel hoppings, the $K_1$ and $K_2$ ($G_1$ and $G_2$) are established for the $y$ and $x$ directions as it is defined in the  Eqn.~\eqref{C4THamiltonian}.
These hopping terms define the momentum-dependent spin interactions that break time-reversal symmetry in the 2D square system.
All hopping amplitudes and energy parameters presented in this work are in units of $eV$.

For the $\mathcal{C}_{4z}\mathbb{T}$-symmetric Hamiltonian given in Eqn.~\eqref{C4THamiltonian}, the energies of the spin-resolved valence band eigenstates are expressed as
\begin{align}
	E^\uparrow_0 &= - \sqrt{ (M^{\uparrow})^2 + |A^{\uparrow}|^2}, \\
	E^\downarrow_0 &= - \sqrt{ (M^{\downarrow})^2 + |A^{\downarrow}|^2},
\end{align}
Notably, along the high-symmetry directions where $k_x = \pm k_y$, one finds $E^\uparrow_0(\mathbf{k}) = E^\downarrow_0(\mathbf{k})$, reflecting the symmetry-enforced degeneracy dictated by the underlying $\mathcal{C}_{4z}\mathbb{T}$ invariance.

Fig. \ref{model} b) presents the spin-resolved band structure calculated for representative parameters: $M_0$=$1$, $K_1$=$2.125$, $K_2 = \tfrac{1}{K_1}$, $G_1$=$1.75$, and $G_2 = \tfrac{1}{G_1}$.  
The energy bands exhibits a momentum-dependent spin splitting induced by orbital hopping asymmetry, which reverses the spin energy band sign between the $\Gamma$–X and $\Gamma$–Y directions.
In addition, a nodal line emerges along the $\Gamma$–M path, indicating symmetry-protected band crossings characteristic of the model \cite{ma2024altermagnetic,nodalines-am}.

In the case that $K_1$=$K_2$, that is when the altermagnetic interaction is due only to the inter-orbital hopping parameters $G_1 \neq G_2$,
we have that $E^\uparrow_0$=$E^\downarrow_0 $ implies the equation $|A^{\uparrow}|$=$|A^{\downarrow}|$,
and this one leads to $\sin(k_x)$=$\pm  \sin (k_y)$. This happens whenever
$k_x$=$\pm k_y$ or $k_x$=$\pi \pm k_y$.

This alternating spin polarization across momentum space is a hallmark of $\mathcal{C}_{4z}\mathbb{T}$-symmetric altermagnetism.

In the case that $G_1=G_2$, the equation $E^\uparrow_0  = E^\downarrow_0 $ implies
that $(M^{\uparrow})^2 - (M^{\downarrow})^2=0$ and this implies that
either $M^{\uparrow} = M^{\downarrow}$ or $M^{\uparrow}+ M^{\downarrow}=0$.
The first equation replicates the equation $k_x= \pm k_y$, while the second
implies the equation
\begin{align} \label{circles change of spin}
\cos(k_x) + \cos(k_y)=\tfrac{2M_0}{K_1+K_2}.
\end{align}
This equation has solutions whenever $|M_0| < |K_1+K_2|$ and this is exactly when
the Hamiltonian $H_4$ has topological insulating features. In Fig. \ref{FeSe_material} c) we see
the spin texture of the upper valence band and the inner ring of change of spin, similar to the one
described the solution of the Eqn. \eqref{circles change of spin}.

\begin{figure}
	\includegraphics[width=8.5cm]{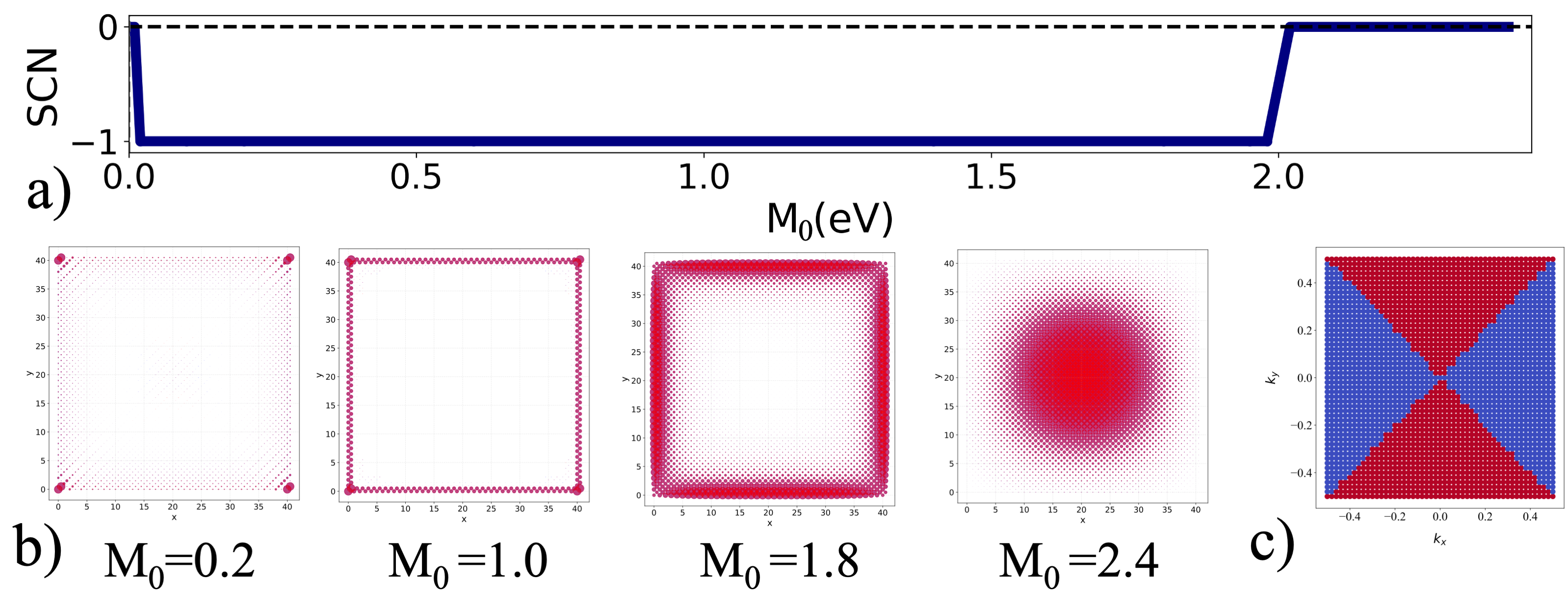}
	\caption{ Hamiltonian $H_4$ of Eqn. \eqref{C4THamiltonian} fixing constants $K_1$=$G_1$=$1.01$ with
	$K_2$=$\tfrac{1}{K_1}$, $G_2$=$\tfrac{1}{G_1}$ and varying $M_0$. a) is the spin$_z$ Chern number
	with respect to $M_0$, b) are the edge states ($|\psi_n(r)|^2$) for $M_0 \in \{0.2,1,1.8,2.4 \}$ and c) is the spin$_z$ texture
	of the upper valence band for the four values of $M_0$ (it is the same texture for the four cases).
	The Hamiltonian models a TI for $0 < |M_0 | < 2$, as the spin Chern number evidences, and the  local charge density of the first three phases exhibit the localization on the boundary.}
	\label{C4Tphase}
\end{figure}

In order to avoid the inner rings  or the extra lines $k_x$=$\pi \pm k_y$ 
in the spin texture of the upper valence band in $H_4$, we need to 
take $K_1 \neq K_2$ and $G_1 \neq G_2$  as in Fig. \ref{C4Tphase} c).

We note here that whenever $K_2=2-K_1$ and $G_1=G_2$ the diagonal term
$M_4^\uparrow$ can be written as the standard term in the BHZ Hamiltonian 
\begin{align}
M:= M_0 - \cos(k_x) - \cos(k_y)
\end{align}
plus a an altermagnetic $J$-term
\begin{align}
J:= (1-K_1) \left(\cos(k_x)-\cos(k_y) \right)
\end{align}
thus having
\begin{align}
M_4^\uparrow = M +J \ \  \mathrm{and} \ \ M_4^\downarrow = M -J.
\end{align}
If the off diagonal term is 
\begin{align}
A= G_1(\sin(k_x) + i \sin(k_y))
\end{align}
the $4\times 4$ Hamiltonian $H_4$ becomes the following matrix:
\begin{align}
\begin{pmatrix}
M + J & A && \\
A^* & -M-J && \\
&& M-J & -A^* \\
&& -A &-M+J  
\end{pmatrix}.
\end{align}

This is the Hamiltonian that has been used in several works that study altermagnetism in Hamiltonians that preserve the  $\mathcal{C}_{4z}\mathbb{T}$ symmetry \cite{ma2024altermagnetic,ma2024quantum,Spin_Chern_Number_Altermagnets}. Whenever $J=0$
this is the original BHZ Hamiltonian \cite{BHZ}.

 The Hamiltonian $H_4$ of \eqref{C4THamiltonian} with $K_1$=$K_2$=$G_1$=$G_2$=$1$ has for spin Chern number
$\mathrm{SCN} = \frac{1}{2}(c_1^\uparrow -c_1^{\downarrow})$ (with $c_1^{\uparrow \downarrow}$ the Chern
number of the spin up and down valence bands)
 the following numbers:
\begin{align} 
\mathrm{SCN} = \left\{ \begin{matrix} \label{spin_Chern_number_C4}
0 & \mathrm{for} & 2 <|M_0|  \\
-1 & \mathrm{for} & 0 < M_0 <2 \\
1 & \mathrm{for}  & -2 < M_0 <0.
\end{matrix} \right.
\end{align}
Therefore we will make the choice of taking
\begin{align} \label{K1_G1}
K_2 = \tfrac{1}{K_1},\ \ G_2 = \tfrac{1}{G_1},
\end{align}
and both $K_1$ and $G_1$  close to $1$.

Figure~\ref{C4Tphase} a) presents the phase diagram of the $\mathcal{C}_{4z}\mathbb{T}$-symmetric Hamiltonian defined in Eqn.~\eqref{C4THamiltonian}, where the SCN is plotted as a function of the $M_0$ parameter (related with the local magnetic moment). 
The SCN was calculated using the computational setup detailed in our previous work \cite{Spin_Weyl_topological_insulators}
Topological phase transition are observed around $M_0$=$2$, where the SCN changes from $-1$ (nontrivial phase) to $0$ (trivial phase), and
around $M_0$=$0$ where the SCN changes from $-1$ to $1$.
Figure~\ref{C4Tphase} b) shows the real-space probability densities of the four edge-states closest to the Fermi level in a cubic geometry for the $M_0$ parameters: $0.2$, $1.0$, $1.8$, and $2.4$. Boundary states were calculated in a square geometry consists of 35$\times$35 unit cells.
It is noted  a transition from strongly localized corner states at $M_0$=$0.2$, to surface-edge states at $M_0$=$1.0$ and $1.8$, and finally to a trivial insulating phase at $M_0$=$2.4$, where the four states nearest the Fermi level become delocalized in the bulk. These edge states are in agreement with previously reported cases \cite{C4T-10.21468/SciPostPhys.15.3.114}.
Figure~\ref{C4Tphase} c) shows the spin texture of the top valence band across the Brillouin zone. This texture is independent of $M_0$, and exhibits the  $\mathcal{C}_{4z}\mathbb{T}$-symmetric shape of the Hamiltonian in Eqn.~\eqref{C4THamiltonian}.
Both, the momentum-space spin texture and the real-space edge-state reflect the $\mathcal{C}_{4z}\mathbb{T}$ symmetry, show the features of a $d$-wave altermagnetic topological phase.


\subsubsection{$\mathcal{C}_{3z} \mathbb{T}$ symmetry}

The case $l$=$3$ requires first the construction of a Hamiltonian which preserves the $(\mathcal{C}_{3z})^2$ symmetry
of Eqn. \eqref{(Cl)2 symmetry}, which in this case it is equivalent to preserving the symmetry $\mathcal{C}_{3z}$ itself.
Define the following terms:
\begin{align} 
M_3^\uparrow =& M_0 - B_0\left(\cos(k_y) +\cos(\tfrac{\sqrt{3}k_x-k_y}{2}) + \cos(\tfrac{-\sqrt{3}k_x-k_y}{2}) \right)\\
J_3^\uparrow =&  J_0 \left( \sin(k_x) \sin(\tfrac{\sqrt{3}k_y-k_x}{2}) \sin(\tfrac{-\sqrt{3}k_y-k_x}{2})\right )  \label{J3}\\
A_3^\uparrow =&A_0  \left(\sin(k_x) +\lambda\sin(\tfrac{\sqrt{3}k_y-k_x}{2}) + \lambda^2\sin(\tfrac{-\sqrt{3}k_y-k_x}{2})   \right) \label{A3}
\end{align}
with  $\lambda = e^{\frac{2 \pi i }{3}}$, and  define the spin up hamiltonian $H_3^\uparrow({\bf{k}})$ 
as follows:
\begin{align}
H_3^\uparrow = \begin{pmatrix}
M_3^\uparrow + J_3^\uparrow & A_3^\uparrow \\
(A_3^\uparrow)^* &   -M_3^\uparrow - J_3^\uparrow 
\end{pmatrix}.
\end{align}

In momentum coordinates the $C_3$ rotation is as follows
\begin{align}
\mathcal{C}_{3z}(k_x,k_y)= (\tfrac{-k_x-\sqrt{3}k_y}{2}, \tfrac{\sqrt{3}k_x-k_y}{2})
\end{align}
and as a representation we associate to $\mathcal{C}_{3z}$ the matrix $\mathrm{diag}( \sigma, \sigma^{-1})$ where $\sigma=e^{\frac{2 \pi i}{6}}$ with $\sigma^2 = \lambda$.

Calculating $A_3^\uparrow(\mathcal{C}_{3z} {\bf{k}})$ we get the equation
\begin{align}
A_3^\uparrow(\mathcal{C}_{3z} {\bf{k}}) = \lambda A_3^\uparrow( {\bf{k}}),
 \end{align}
which implies the matrix equation
\begin{align}
\mathrm{diag}( \sigma, \sigma^{-1}) H^\uparrow_3({\bf{k}}) \mathrm{diag}( \sigma^{-1}, \sigma)
= H_3^\uparrow(\mathcal{C}_{3z} {\bf{k}}) 
\end{align}
thus implying Eqn. \eqref{(Cl)2 symmetry} when applied twice.

The Hamiltonian $H_3^\downarrow({\bf{k}})$ is defined following Eqn. \eqref{definition H(ClT)}  
\begin{align}
H_3^\downarrow({\bf{k}}) := \mathrm{diag}( \sigma, \sigma^{-1}) H^\uparrow_3(-\mathcal{C}_{3z}{\bf{k}}) \mathrm{diag}( \sigma^{-1}, \sigma)
\end{align}
implying that
\begin{align}
M_3^\downarrow({\bf{k}}) :=& M_3^\uparrow(-\mathcal{C}_{3z}{\bf{k}}) = M_3^\uparrow({\bf{k}}), \\
J_3^\downarrow({\bf{k}}) := &J_3^\uparrow(-\mathcal{C}_{3z}{\bf{k}})  = -J_3^\uparrow({\bf{k}}),\\
A_3^\downarrow({\bf{k}}) := & \lambda A_3^\uparrow(-\mathcal{C}_{3z}{\bf{k}})^*  = -\lambda(\lambda A_3^\uparrow({\bf{k}}))^*=-A_3^\uparrow({\bf{k}})^*.
\end{align}

The $4 \times 4$ Hamiltonian $H_3$ written in terms of $M^\uparrow$, $J^\uparrow$ and $A^\uparrow$
is as follows:
\begin{align}\label{C3THamiltonian}
H_3({\bf{k}})= 
\begin{pmatrix}
M_3^\uparrow+ J_3^\uparrow & A_3^\uparrow & &\\
(A_3^\uparrow)^* & -M_3^\uparrow - J_3^\uparrow &  &\\
&  & M_3^\uparrow- J^\uparrow & -(A_3^\uparrow)^* \\
 & &  -A_3^\uparrow & -M_3^\uparrow+ J_3^\uparrow 
\end{pmatrix}.
\end{align}

\begin{figure}
	\includegraphics[width=8.5cm]{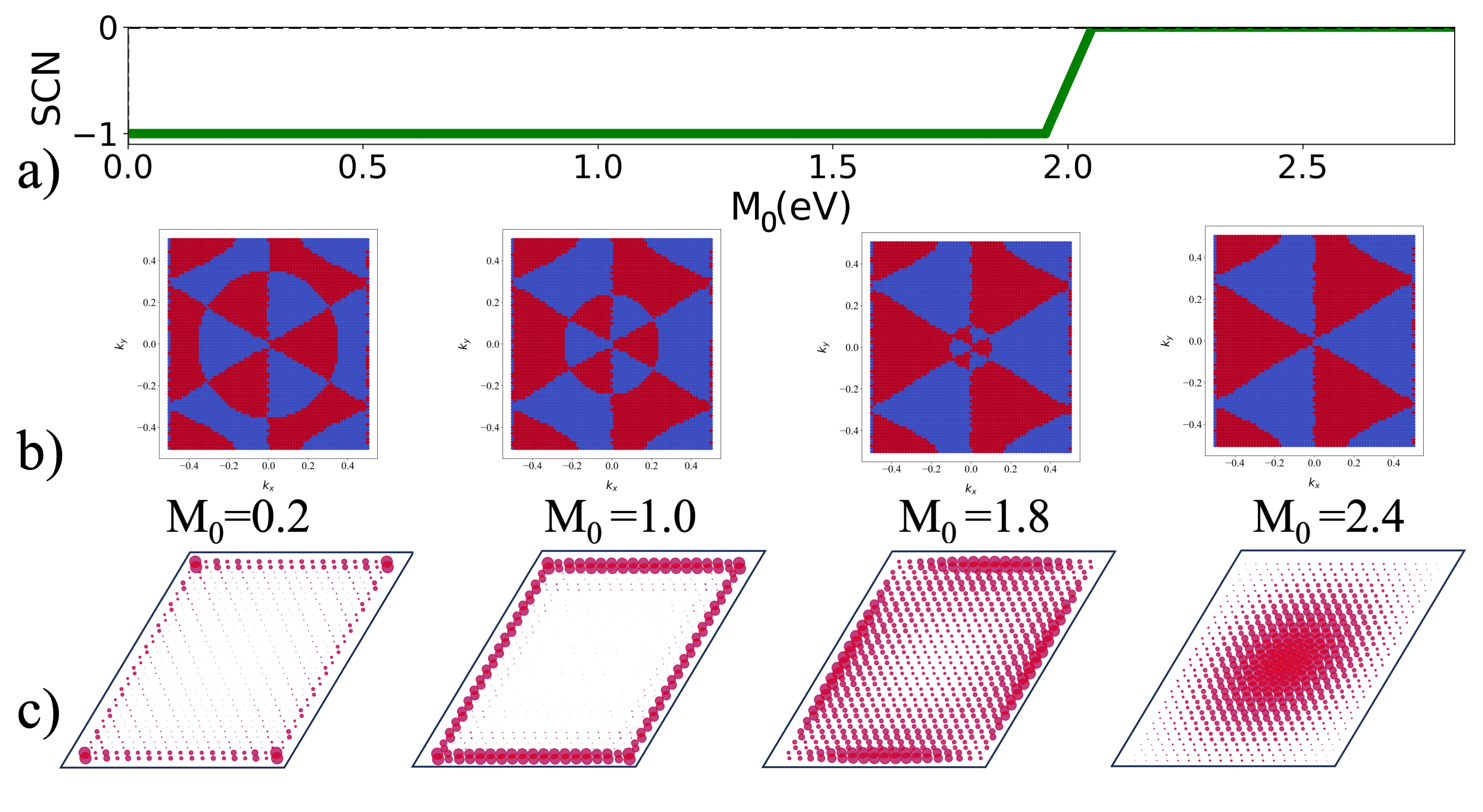}
   \caption{(a) Phase diagram of the $\mathcal{C}_{3z}\mathbb{T}$-symmetric altermagnetic model described by Eqn.~\eqref{C3THamiltonian}, showing the SCN as a function of the mass term $M_0$. A topological phase transition occurs near $M_0$=$2$.  
	(b) Spin-projected states of the top valence band across the Brillouin zone, and  
	(c) Spin-resolved edge states for a finite system, calculated at $M_0=0.2$, $1.0$, $1.8$, and $2.4$, with fixed parameters $A_0$=$1.01$, $B_0$=$2/3$ and $J_0$=$0.01$.  
	The spin textures exhibit an $f$-wave AM type characteristic of the $\mathcal{C}_{3z}\mathbb{T}$ symmetry, with a nodal ring whose radius decreases as $M_0$ increases. Both the spin textures and edge-state distributions manifest the $\mathcal{C}_{3z}\mathbb{T}$ symmetry of the Hamiltonian in Eqn.~\eqref{C3THamiltonian}.}

	\label{C3Tphase}
\end{figure}

The energies of the valence spin up and spin down eigenstates are respectively
\begin{align}
E^\uparrow_0 &= - \sqrt{ (M_3^{\uparrow}+J_3^\uparrow)^2 + |A_3^{\uparrow}|^2}, \\
E^\downarrow_0 &=  - \sqrt{ (M_3^{\uparrow}-J_3^\uparrow)^2 + |A_3^{\uparrow}|^2}
\end{align}
and the energies are equal whenever $M^{\uparrow}$=$0$ or $J^\uparrow$=$0$. 
The second equation implies that $k_x$=$0$ or $k_x$=$\pm \sqrt{3} k_y$, which is 
a triangular tesselation of the momentum space,  and the first
implies the equation
\begin{align} \label{nodal-ring-eqn}
\cos(k_y) +\cos(\tfrac{\sqrt{3}k_x-k_y}{2}) + \cos(\tfrac{-\sqrt{3}k_x-k_y}{2}) = \tfrac{M_0}{B_0}
\end{align}
which has solutions only when $\big| \tfrac{M_0}{B_0} \big|< 3$. Since $H_3$ is precisely
TI whenever $\big| \tfrac{M_0}{B_0} \big|< 3$, we see that the appearance of a nodal line
in the Hamiltonian is unavoidable for the TI phase. This nodal line looks like a circle around the $\Gamma$ point
and can be appreciated in Fig. \ref{C3Tphase} b) where it induces a change of spin the texture around $\Gamma$.

Since we can scale all constants of the Hamiltonian at the same time, we will choose
\begin{align}
B_0= \tfrac{2}{3}
\end{align}
for the relevant values of $M_0$ be the same as in the case of the Hamiltonian $H_4$ of Eqn. \eqref{C4THamiltonian}.
The constants $J_0$ and $A_0$ will be close to $0$ and $1$ respectively, and
whenever $J_0$=$0$ and $A_0$=$1$ the Hamiltonian $H_3$  is gapless for $M_0 \in \{-2,0,2\}$
and the SCN of the valence bands is 
\begin{align} \label{spin_Chern_number_C3}
\mathrm{SCN} = \left\{ \begin{matrix}
0 & \mathrm{for} & 2 < |M_0| \\
-1 & \mathrm{for} & 0 < |M_0| <2.
\end{matrix} \right.
\end{align}
Note that the SCN is only $-1$ because the Hamiltonian $H_3^\uparrow$ has Chern number $-1$ for $0 < |M_0|<2$.
This case differs from the Hamiltonian $H_4$ where the SCN changes from $-1$ to $1$ on $M_0$=$0$ as it is shown
in Eqn. \eqref{spin_Chern_number_C4} and illustrated in Fig. \ref{C4Tphase}.

Figure \ref{C3Tphase} a) illustrates the topological phase transition in the  $\mathcal{C}_{3z}\mathbb{T}$-symmetric model described by Eqn.  ~\eqref{C3THamiltonian}, controlled by the onsite mass term $M_0$. 
A transition from a nontrivial ($\mathrm{SCN}$=$1$) to a trivial ($\mathrm{SCN}$=$0$) phase occurs near $M_0$=$2$. This transition highlights the tunability of the topological phase with the compensate local magnetization of the opposite atomic sublattices, described in the $M_0$ value. 
The quantized $\mathrm{SCN}$=$1$ reflects a robust topological invariant \cite{Prodan-SCN}, in agreement with the topological classification of 2D-systems preserving combined $\mathcal{C}_{3z}\mathbb{T}$ symmetry.

Figure \ref{C3Tphase} b) and \ref{C3Tphase} c) display the spin texture on the top valence band across the Brillouin zone and the corresponding spin-resolved edge-state spectra for a finite sample, respectively. Boundary states were calculated in an geometry composed of 35$\times$35 unit cells.
These results are computed for $M_0$ = $0.2$, $1.0$, $1.8$, and $2.4$, with fixed parameters $A_0$ = $1.01$ and $J_0$ = $0.01$, representing inter-sublattice and intra-sublattice hopping amplitudes, respectively. 
In all cases, the spin textures reveal an $f$-wave altermagnetic type, which is characteristic of the $\mathcal{C}_{3z}\mathbb{T}$-protected symmetry \cite{Smejkal2022}.
A nodal ring centered at the $\Gamma$ point where the spin texture reverses, is observed, as predicted by the Eqn. \eqref{nodal-ring-eqn}.
Its radius progressively decreases with increasing $M_0$.
The presence of corner and edge modes within the bulk energy gap confirms the nontrivial topology for $M_0 < 2$, while for $M_0$=$2.4$, the absence of edge states is an signal a trivial insulating phase.


 \subsubsection{Topological indicator} 
  
  The Bloch bundle of the valence bands for the gapped Hamiltonians $H_4$ and $H_3$ of Eqns. \eqref{C4THamiltonian}
  and \eqref{C3THamiltonian}, define elements in the Magnetic Equivariant  K-theory groups \cite{MEK} of the 2D torus 
  $\mathcal{K}_{\mathcal{C}_{4z}\mathbb{T}}^0(T^2)$ and  $\mathcal{K}_{\mathcal{C}_{3z}\mathbb{T}}^0(T^2)$
  respectively. The bulk invariant of  the K-theory group $\mathcal{K}_{\mathcal{C}_{4z}\mathbb{T}}^0(T^2)$ was shown to be $\mathbb{Z}/2$ and it was further shown that it could be measured with the value mod $2$ of the SCN\cite{Spin_Chern_Number_Altermagnets}. 
  Similarly, the bulk invariant of the K-theory group $\mathcal{K}_{\mathcal{C}_{3z}\mathbb{T}}^0(T^2)$ is also $\mathbb{Z}/2$ and can be measured with the SCN.
  
  Hence the SCN on the gapped Hamiltonians provides the topological bulk invariant of the system. The SCN is invariant
  while the altermagnet Hamiltonians remain gapped and the symmetry $\mathcal{C}_{l}\mathbb{T}$ is not broken.
  
\subsection{3D Hamiltonians}

The Hamiltonian in Eqn. \eqref{2D Hamiltonian} can be extended to three dimensions by incorporating an intrasublattice hopping along the $z$-axis 
and an off-diagonal $k_z$-dependent term that preserves the $\mathcal{C}_l \mathbb{T}$ symmetry and maintains the bulk energy gap.

One proposal for such a Hamitlonian is the following:
\begin{align}
 \mathcal{H}(k_x,k_y,k_z) = \begin{pmatrix}
 \mathcal{H}^{\uparrow}(k_x,k_y,k_z)  & D_0 \sin(k_z) \sigma_x \\
  D_0 \sin(k_z) \sigma_x& \mathcal{H}^{\downarrow}(k_x,k_y,k_z) 
 \end{pmatrix}
 \end{align}
where the up and down  sectors of the Hamiltonian are defines as follows:
\begin{align}
 \mathcal{H}^{\uparrow}(k_x,k_y,k_z) : =  H^{\uparrow}(k_x,k_y) - \cos(k_z)\sigma_z \\
 \mathcal{H}^{\downarrow}(k_x,k_y,k_z) : =  H^{\downarrow}(k_x,k_y)- \cos(k_z)\sigma_z 
\end{align}
 with $H^{\uparrow}(k_x,k_y) $ and $H^\downarrow(k_x,k_y)$  the Hamiltonians defined in Eqns. \eqref{definition H up}
and \eqref{definition H(ClT)} respectively, $\sigma_x$ and $\sigma_y$ are the Pauli matrices and $D_0$ is a non-zero real parameter.

This Hamiltonian preserves the symmetry $\mathcal{C}_{lz} \mathbb{T}$ where $\mathcal{C}_{lz} (k_x,k_y,k_z)=(\mathcal{C}_l(k_x,k_y),k_z)$ and the matrix associated to $\mathcal{C}_{lz} \mathbb{T}$ is
\begin{align}
\begin{pmatrix}
\mathcal{C}_l & 0 \\
0&  \mathcal{C}_l^{-1}  
 \end{pmatrix} 
  \begin{pmatrix}
0 & \mathrm{I}_{2} \\
- \mathrm{I}_2 &  0  
 \end{pmatrix} \mathbb{K}.
\end{align}
The Hamiltonian is gapped as long as the Hamiltonian $\mathcal{H}^{\uparrow}(k_x,k_y,k_z) $ remains gapped
for $k_z=0$ and $k_z=\pi$.

\subsubsection{$\mathcal{C}_{4z} \mathbb{T}$ symmetry}

The spin up and spin down sectors of the Hamiltonian become
\begin{align} \label{definition H up 3D C4T}
\mathcal{H}_4^{\uparrow \downarrow}({\bf{k}}) =
\begin{pmatrix}
\mathcal{M}_4^{\uparrow \downarrow}({\bf{k}}) & A_4^{\uparrow \downarrow}({\bf{k}})  \\
A_4^{\uparrow \downarrow}({\bf{k}})^* & -\mathcal{M}_4^{\uparrow \downarrow}({\bf{k}}) 
\end{pmatrix}
 \end{align} 
where
 \begin{align} 
 \mathcal{M}_4^{\uparrow}({\bf{k}}) & = M_0 - K_1 \cos (k_x) - K_2\cos(k_y)- \cos(k_z),\\
 A_4^{\uparrow}({\bf{k}})  & = G_1 \sin(k_x) + i G_2 \sin (k_y),\\
 \mathcal{M}_4^{\downarrow}({\bf{k}}) & = M_0 - K_2\cos(k_x) - K_1 \cos (k_y) -\cos(k_z),\\
 A_4^{\downarrow}({\bf{k}})  &  =-G_2 \sin(k_x) + i G_1 \sin (k_y).
 \end{align}

The 3D Hamiltonian is thus
\begin{align} \label{3D_H4_Hamiltonian}
 \mathcal{H}_4(k_x,k_y,k_z) = \begin{pmatrix}
 \mathcal{H}_4^{\uparrow}(k_x,k_y,k_z)  & D_0 \sin(k_z) \sigma_x \\
  D_0 \sin(k_z) \sigma_x& \mathcal{H}_4^{\downarrow}(k_x,k_y,k_z) 
 \end{pmatrix},
 \end{align}
it preserves the symmetry $\mathcal{C}_{4z} \mathbb{T}$, breaks time reversal
as defined in Eqn. \eqref{T_and_C_l}, preserves inversion symmetry and it is gapped as long as $D_0 \neq 0$.

Whenever $G_1$=$G_2$=$K_1$=$K_2$=$1$ the system looses its altermagnetic feature
and is trivial for $|M_0| > 3$, strong TI for $1 < |M_0| < 3 $ with two spin Weyl points in the line $k_x$=$0$=$k_y$
when $M_0$ is positive and  two spin Weyl points in the line $k_x$=$\pi$=$k_y$ when $M_0$ is negative,
 and weak TI for $|M_0 | <1$ with four spin Weyl points, two in the line $k_x$=$0$, $k_y$=$\pi$ 
 and two in the line $k_x$=$\pi$, $k_y$=$0$. This feature is similar to the one appearing in the 3D BHZ Hamiltonian
  \cite{Spin_Weyl_topological_insulators}.
   In terms of the SCN on the planes $k_z$=$0$ and $k_z$=$\pi$ the information has been summarized in Table \ref{table_SCN_C4T}.
 \begin{table}[!h]
  \begin{center}
\begin{tabular}{ c|ccccc } 
\hline
 $M_0$ & $(-\infty,-3)$ & $(-3,-1)$ & $(-1,1)$ & $(1,3)$ & $(3, \infty)$\\ 
 \hline
 SCN$(k_z$=$\pi)$ & $0$ &$1$ &$-1$ & $0$ & $0$\\ 
  SCN$(k_z$=$0)$ & $0$ & $0$ & $1$& $-1$& $0$ \\ 
 \hline
  Top. type& Tr. & STI & WTI & STI & Tr.\\
  \hline
\end{tabular}
\end{center}
\caption{SCN on the planes $k_z$=$\pi$ and $k_z$=$0$ of the Hamiltonian $\mathcal{H}_4$ for different values of $M_0$
and $G_1$=$G_2$=$K_1$=$K_2$=$1$ and $D_0$=$0.1$. The last row indicates its topological nature: trivial insulator (Tr.), strong TI (STI) or weak TI (WTI).} 
 \label{table_SCN_C4T}
 \end{table}

 \begin{figure} 
	\includegraphics[width=8.7cm]{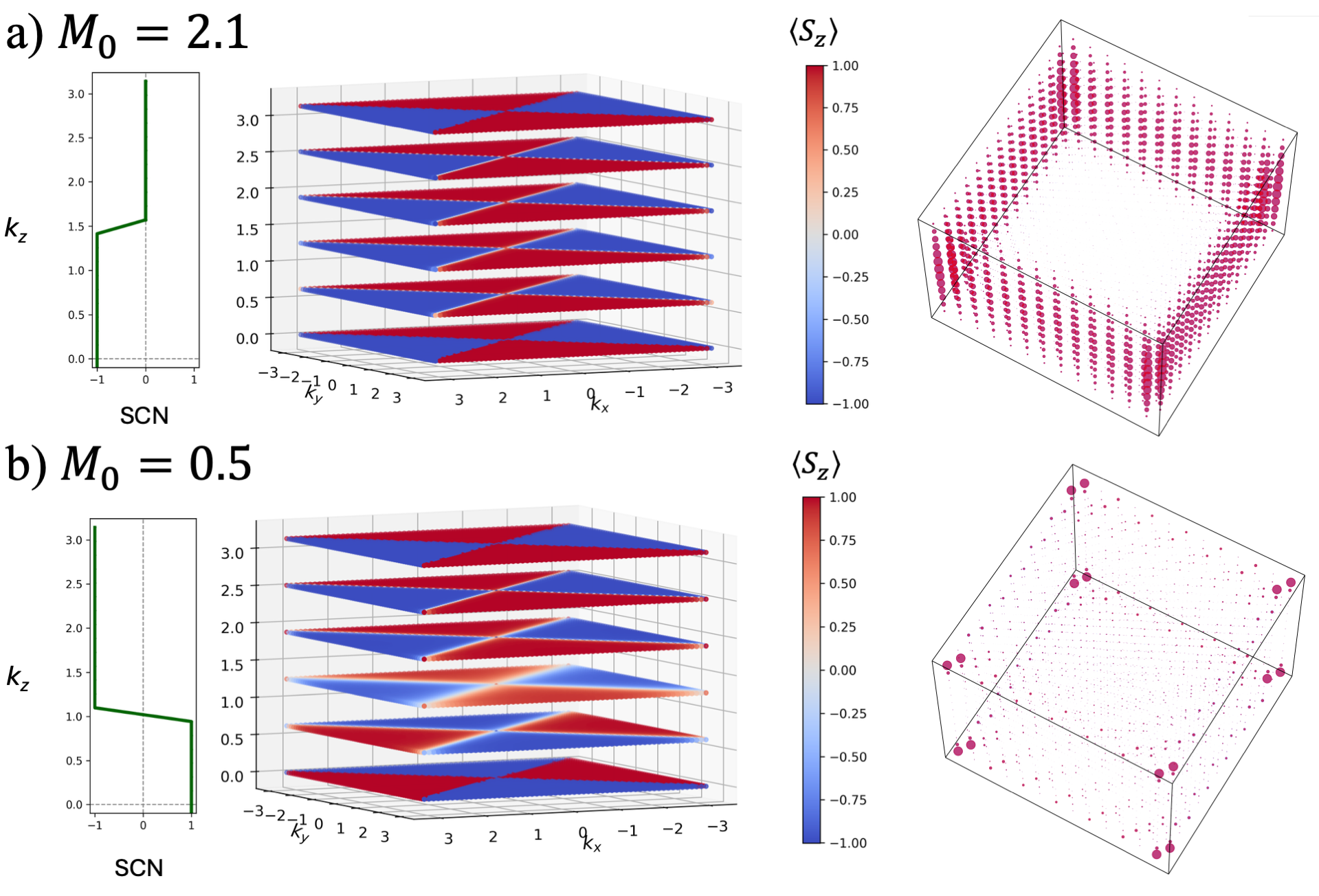}
	\caption{ $\mathcal{H}_4({\bf{k}})$ Hamiltonian of Eqn. \eqref{3D_H4_Hamiltonian} with  $K_1$=$1.01$, $G_1$=$1.01$,  $D_0$=$0.1$, a) $M_0$=$2.1$ and b) $M_0$=$0.5$.  The first graph is the SCN of the valence bands on planes perpendicular to the $k_z$-axis. The second
	is the spin$_z$ texture of the upper valence band on the BZ. The third are the conducting boundary states.
	a) is a strong TI because of the change of SCN mod 2. The SCN changes by 1 where the spin Weyl point is located, on the $k_z$-axis
	with $(k_x,k_y)$=$(0,0)$.
	The altermagnetic feature is almost constant in all value of $k_z$. The edge states are localized on the faces parallel to the $z$-axis. b) is a weak TI because the SCN mod 2 is always 1. The SCN changes from 1 to -1 where the spin Weyl points are located, one on the $k_z$-axis with $(k_x,k_y)$=$(0,\pi)$, and the other of the $k_z$-axis with $(k_x,k_y)$=$(\pi,0)$,. The spin texture  flips on the $k_z$ locations of the spin Weyl points. The boundary states are localized in the corners, thus inducing a HOTI phase. 
	The boundary state calculations were performed on a 13$\times$13$\times$13 finite cubic geometry.
}
	\label{C4T-figure}
\end{figure}
 
The parameters  $K_2$ and $G_2$ will be chosen as in Eqn. \eqref{K1_G1} and $K_1$ and $G_1$ will
be values  close to  $1$. 

In Fig.\ref{C4T-figure}, we present the SCN and  the spin$_z$ texture of the upper valence band as a function of $k_z$, and the four boundary states of the 3D Hamiltonian $\mathcal{H}_4({\bf{k}})$ defined in Eqn.\eqref{3D_H4_Hamiltonian}.
Here we have fixed
 $K_1$=$1.01$=$G_1$ and $D_0$=$0.1$ we have plotted the cases  a) $M_0$=$2.1$ and b) $M_0$=$0.5$.
The case a) is a strong TI with constant spin texture on all $k_z$ planes, two spin Weyl points and
change of SCN on the planes $k_z$=$0$ and $k_z$=$\pi$. The boundary states are localized on the faces
perpendicular to the $z$-axis. The case b) is a weak TI
with same SCN mod $2$ on the planes $k_z$=$0$ and $k_z$=$\pi$, four spin Weyl points located the $k_z$-axis with
$(k_x,k_y)$=$(0,\pi)$ and $(k_x,k_y)$=$(\pi,0)$, and spin texture which flips on the $k_z$ plane where the spin Weyl points are located. 
The boundary states near the Fermi level are spatially localized at the corners of the finite cubic geometry, indicating the presence of a HOTI phase. 
This behavior is analogous to the corner modes protected by $\mathcal{C}_{4z}\mathbb{T}$ symmetry model reported by Schindler \textit{et al.} \cite{HOTI}


 The results presented in Fig.~\ref{C4T-figure} show the interplay between symmetry-protected topology and spatial localization of boundary states in a $\mathcal{C}_{4z}\mathbb{T}$-symmetric system. 
As the mass parameter $M_0$ is tuned, the wavefunction states of the Hamiltonian of Eqn. \eqref{3D_H4_Hamiltonian} undergoes topological phase transitions that manifest not only through quantized spin Chern numbers but also through the real-space edge-states distribution.
The corner-localized to surface states, for $M_0$ close to 0, reflects the bulk boundary correspondences of a HOTI system enforced by the  $\mathcal{C}_{4z}\mathbb{T}$ symmetry. This behavior is consistent with previous HOTI systems with combined four-fold rotation and time-reversal symmetry\cite{ojito2024cframeworkhigherorderbulkboundarycorrespondences,HOTI-AMs}.
In addition, the surface states for $M_0$ close to 2 are more extended to the surface showing a strong topological insulator character, which is confirmed with the change of SCN from 0 to 1.

\subsubsection{$\mathcal{C}_{3z} \mathbb{T}$ symmetry}

The 3D Hamiltonian coming from the 2D Hamiltonian of Eqn. \eqref{C3THamiltonian} is:
\begin{align}\label{C3T-3D-Hamiltonian}
\mathcal{H}_3({\bf{k}})= 
\begin{pmatrix}
\mathcal{M}_3^\uparrow+ J_3^\uparrow & A_3^\uparrow & & D_0 \sin(k_z)\\
(A_3^\uparrow)^* & -\mathcal{M}_3^\uparrow - J_3^\uparrow & D_0 \sin(k_z)  &\\
&  D_0 \sin(k_z)& \mathcal{M}_3^\uparrow- J^\uparrow & -(A_3^\uparrow)^* \\
D_0 \sin(k_z) & &  -A_3^\uparrow & -\mathcal{M}_3^\uparrow+ J_3^\uparrow 
\end{pmatrix}
\end{align}
where 
\begin{align} 
\mathcal{M}_3^\uparrow = M_0 &- \cos(k_z) -\\ & B_0 \left(\cos(k_y) +\cos(\tfrac{\sqrt{3}k_x-k_y}{2}) + \cos(\tfrac{-\sqrt{3}k_x-k_y}{2})\right)  \nonumber
\end{align}
and $J_3^\uparrow$ and $A_3^\uparrow$ remain as in Eqns. \eqref{J3} and \eqref{A3}.

This Hamiltonian preserves the $\mathcal{C}_{3z} \mathbb{T}$ symmetry together with time reversal and
breaks inversion symmetry.
Whenever $J_0$=$0$ and $A_0$=$1$, this Hamiltonian is gapped as long as $M_0\notin\{-3,-1,1,3\}$ and $D_0 \neq 0$.
The SCN on the planes $k_z$=$0$ and $k_z$=$\pi$ are as the ones of Table \ref{table_SCN_C3T} and 
the value of $M_0$ determines its topological features.
The altermagnetic flavour of the system is obtained when $J_0 \neq 0$. 

\begin{table}[!h]
  \begin{center}
\begin{tabular}{ c|ccccc } 
\hline
 $M_0$ & $(-\infty,-3)$ & $(-3,-1)$ & $(-1,1)$ & $(1,3)$ & $(3, \infty)$\\ 
 \hline
 SCN$(k_z$=$\pi)$ & $0$ &$1$ &$1$ & $0$ & $0$\\ 
  SCN$(k_z$=$0)$ & $0$ & $0$ & $1$& $1$& $0$ \\ 
 \hline
  Top. type& Tr. & STI & WTI & STI & Tr.\\
  \hline
\end{tabular}
\end{center}
\caption{SCN on the planes $k_z$=$\pi$ and $k_z$=$0$ of the Hamiltonian $\mathcal{H}_3$ for different values of $M_0$
with $J_0$=$0$, $A_0$=$1$ and $D_0$=$0.1$. The last row indicates its topological nature: trivial insulator, strong TI or weak TI.} 
 \label{table_SCN_C3T}
 \end{table}

\begin{figure} 
	\includegraphics[width=8.5cm]{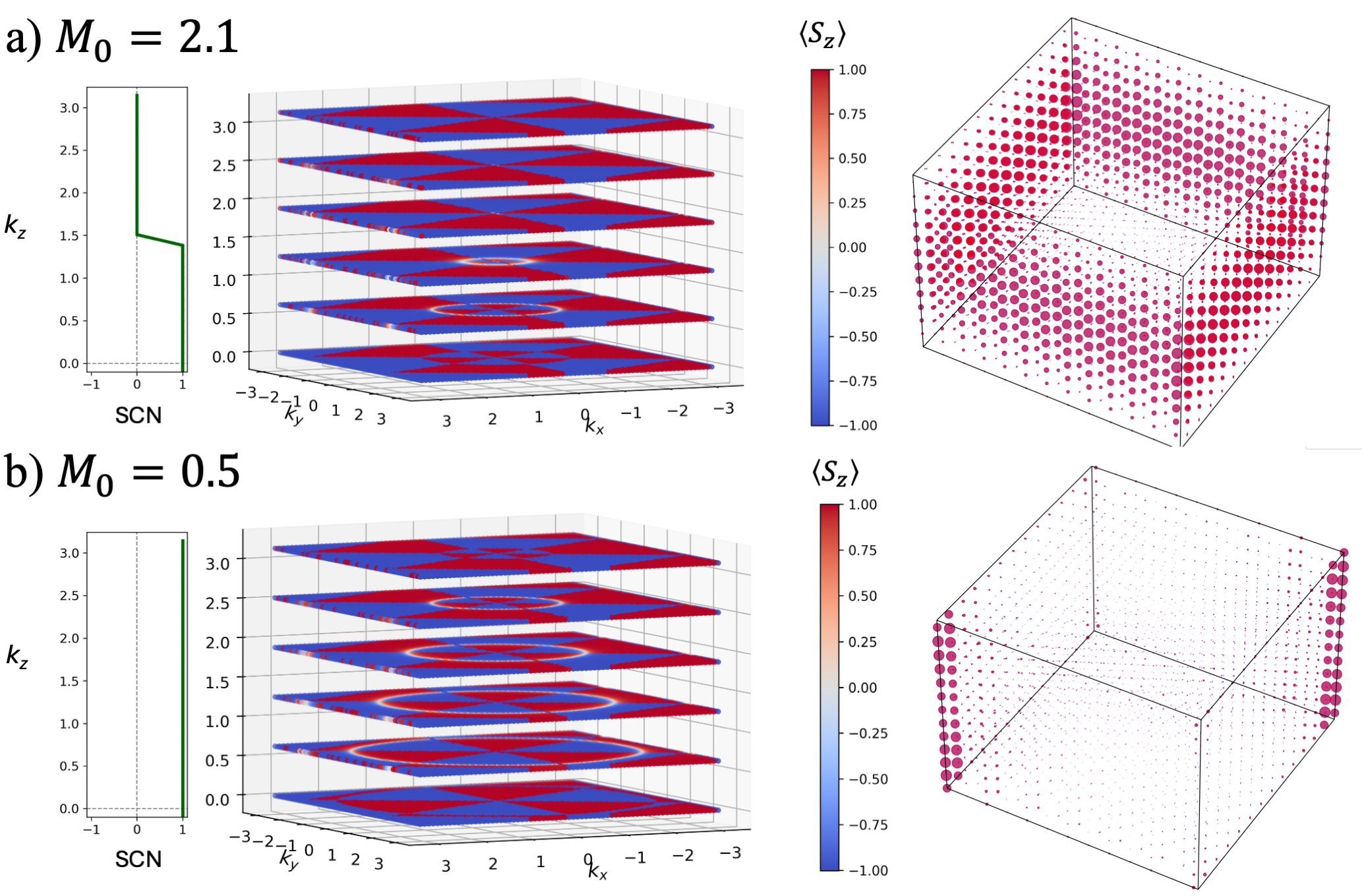}
	\caption{ $\mathcal{H}_3({\bf{k}})$ Hamiltonian of Eqn. \eqref{C3T-3D-Hamiltonian} with constants 
	 $B_0$=$2/3$, $J_0$=$0.01$, $A_0$=$1.01$, $D_0$=$0.1$, a) $M_0$=$2.1$ and b) $M_0$=$0.5$. 
	  The first graph is the SCN of the valence bands on planes perpendicular to the $k_z$-axis. The second
	is the spin$_z$ texture of the upper valence band on the BZ. The third are the conducting boundary states.
	a) is a strong TI because of the change of SCN mod 2. The SCN changes by 1 where the spin Weyl point is located, on the $k_z$-axis
	with $(k_x,k_y)$=$(0,0)$.
	The altermagnetic feature is similar for all value of $k_z$, except when the presence of a nodal sphere
	makes the the spin texture to change of sign.
	The edge states are localized on the faces parallel to the $z$-axis. b) is a weak TI because the SCN is always 1. 
	There is no change of SCN since there are no spin Weyl points. The spin texture flips sign from $k_z$=$\pi$ to $k_z$=$0$
	due to the nodal sphere present on the whole BZ.
	 The boundary states are localized in the hinges, thus inducing a HOTI phase.
    Boundary states were computed using a finite cubic system of size 13$\times$13$\times$13.
	}
	\label{3D-C3T}
\end{figure}

 We have plotted in Fig. \ref{3D-C3T} the SCN with respect to $k_z$, the spin texture and the edge states
 for a) $M_0$=$2.1$ and b) $M_0$=$0.5$, fixing
$B_0$=$2/3$, $J_0$=$0.01$, $A_0$=$1.01$ and $D_0$=$0.1$. The case
 a) is a strong TI with change of SCN from $0$ in $k_z$=$\pi$ to $1$ in $k_z$=$0$,
 spin Weyl points localized in the $k_z$-axis with $(k_x,k_y)$=$(0,0)$, edge states localized in the planes parallel to the $z$-axis, and whose spin texture changes in the presence of a nodal sphere. 
 In contrast, case (b) corresponds to a weak TI phase, where the SCN remains constant across $k_z$, and no spin Weyl points are present. 
 The system instead hosts hinge states propagating along the $z$-axis, accompanied by a sign reversal of the spin texture between $k_z$=$0$ and $k_z$=$\pi$.
 Additionally, a symmetry-protected nodal line is observed in the spin texture across varying $k_z$ planes.
 These results, together with the recent discovery of floating edge bands in altermagnetic BHZ-type systems \cite{floatingBHZ-AM}, highlight the unconventional boundary phenomena that can arise in altermagnetic topological phases.


\subsubsection{Topological invariants}

The 3D bulk invariants of the $\mathcal{C}_{4z} \mathbb{T}$ symmetry can be determined
from the Magnetic Equivariant K-theory\cite{MEK} groups   $\mathcal{K}_{\mathcal{C}_{4z}\mathbb{T}}^0(T^3)$
of the 3D torus $T^3$ (similarly for $\mathcal{C}_{3z} \mathbb{T}$). 

To calculate this K-theory group we are going to use the Mayer-Vietoris sequence\cite{MEK} and the information
we already know from the 2D case\cite{Spin_Chern_Number_Altermagnets}.  
Cover $T^3$ with open sets $U$ and $V$ where $U = T^2 \times (-\tfrac{3\pi}{4},  \tfrac{3\pi}{4})$
and $V= T^2 \times  (\tfrac{\pi}{4},\tfrac{7\pi}{4})$. The intersection $U \cap V$ is equivaraintly homotopic
to the union of the planes $k_z$=$\tfrac{\pi}{2}$ and  $k_z$=$-\tfrac{\pi}{2}$,  $U$ and $V$ are equivaraintly
homotopic to $k_z$=$0$ and  $k_z$=$\pi$ respectively. Therefore we have the following
isomorphisms:
\begin{align}
\mathcal{K}_{\mathcal{C}_{4z}\mathbb{T}}^0(U)  \cong \mathcal{K}_{\mathcal{C}_{4z}\mathbb{T}}^0(V) \cong
\mathcal{K}_{\mathcal{C}_{4z}\mathbb{T}}^0(T^2)\\
\mathcal{K}_{\mathcal{C}_{4z}\mathbb{T}}^0(U \cap V) \cong K_{\mathcal{C}_{2z}}^0(T^2)
\end{align}
where $K_{\mathcal{C}_{2z}}^0(T^2)$ stands for the equivariant complext K-theory of the 2-fold rotation $\mathcal{C}_{2z}$
on $T^2$.

Since we know that $K_{\mathcal{C}_{2z}}^{-1}(T^2)=0$,\cite{MEK}  and that the restriction map
\begin{align}
r:\mathcal{K}_{\mathcal{C}_{4z}\mathbb{T}}^0(T^2) \to K_{\mathcal{C}_{2z}}^0(T^2)
\end{align}
is injective on its torsion free part\cite{SUX}, we can deduce that we have a exact sequene
\begin{align}
0 \to \mathcal{K}_{\mathcal{C}_{4z}\mathbb{T}}^0(T^3)  \to \mathcal{K}_{\mathcal{C}_{4z}\mathbb{T}}^0(T^2)^{\oplus 2} \to
K_{\mathcal{C}_{2z}}^0(T^2)
\end{align}
where the right hand side map is $a_0 \oplus a_1 \mapsto r(a_0) + r(a_1).$
We know that\cite{Spin_Chern_Number_Altermagnets}
\begin{align}
\mathcal{K}_{\mathcal{C}_{4z}\mathbb{T}}^0(T^2) \cong \mathbb{Z}^{\oplus 2} \oplus \mathbb{Z}/2
\end{align}
where the $\mathbb{Z}/2$ invariant can be obtained from the parity of the SCN. Hence we see that
our desired K-theory group is:
\begin{align}
\mathcal{K}_{\mathcal{C}_{4z}\mathbb{T}}^0(T^3)  \cong \mathbb{Z}^{\oplus 2} \oplus \mathbb{Z}/2^{\oplus 2},
\end{align}
 where the bulk invariants are precisely the two copies of $\mathbb{Z}/2$, and are measured calculating
 the SCN on the planes $k_z$=$0$ and $k_z$=$\pi$.
 
 The torsion subgroup of $\mathcal{K}_{\mathcal{C}_{4z}\mathbb{T}}^0(T^3)  $ which is the two copies of 
 $\mathbb{Z}/2$ is the TI indicator for this symmetry. Exactly the same applies to $\mathcal{K}_{\mathcal{C}_{3z}\mathbb{T}}^0(T^3)  $.
 
Therefore  the four different phases that can appear 
 are presented in Table \ref{table_3D-top-inv} and are compatible with the information provided by the 
 SCN presented in Tables \ref{table_SCN_C4T} and \ref{table_SCN_C3T}.
 
 \begin{table}[!h]
  \begin{center}
\begin{tabular}{ c|cccc }
\hline 
 Top. type & Tr. & STI & WTI & STI  \\
 \hline
 SCN$(k_z$=$\pi)$ & $0$ &$1$ &$1$ & $0$\\ 
  SCN$(k_z$=$0)$ & $0$ & $0$ & $1$& $1$\\ 
 \hline
\end{tabular}
\end{center}
\caption{Indicators for the topological invariants appearing in the torsion subgroup of $\mathcal{K}_{\mathcal{C}_{4z}\mathbb{T}}^0(T^3)$ and  $\mathcal{K}_{\mathcal{C}_{3z}\mathbb{T}}^0(T^3)$. Here the torsion subgroup consists of two copies of $\mathbb{Z}/2$, the four possibilities
for the invariant apear in the table and they are obtained from the calculation of the SCN mod $2$ on the planes
fixed by $\mathcal{C}_{4z}\mathbb{T}$ or $\mathcal{C}_{3z}\mathbb{T}$ which are $k_z$=$0$ and $k_z$=$\pi$. If the SCN is $1$ and it does not change then
we are in the weak TI realm, while the change of the SCN detects the strong TI feature. The
topological nature can be obtained from the information of the SCN as presented in Tables \ref{table_SCN_C4T} and \ref{table_SCN_C3T}.} 
 \label{table_3D-top-inv}
 \end{table}

\section{Material realization}

\begin{figure}
	\includegraphics[width=8.5cm]{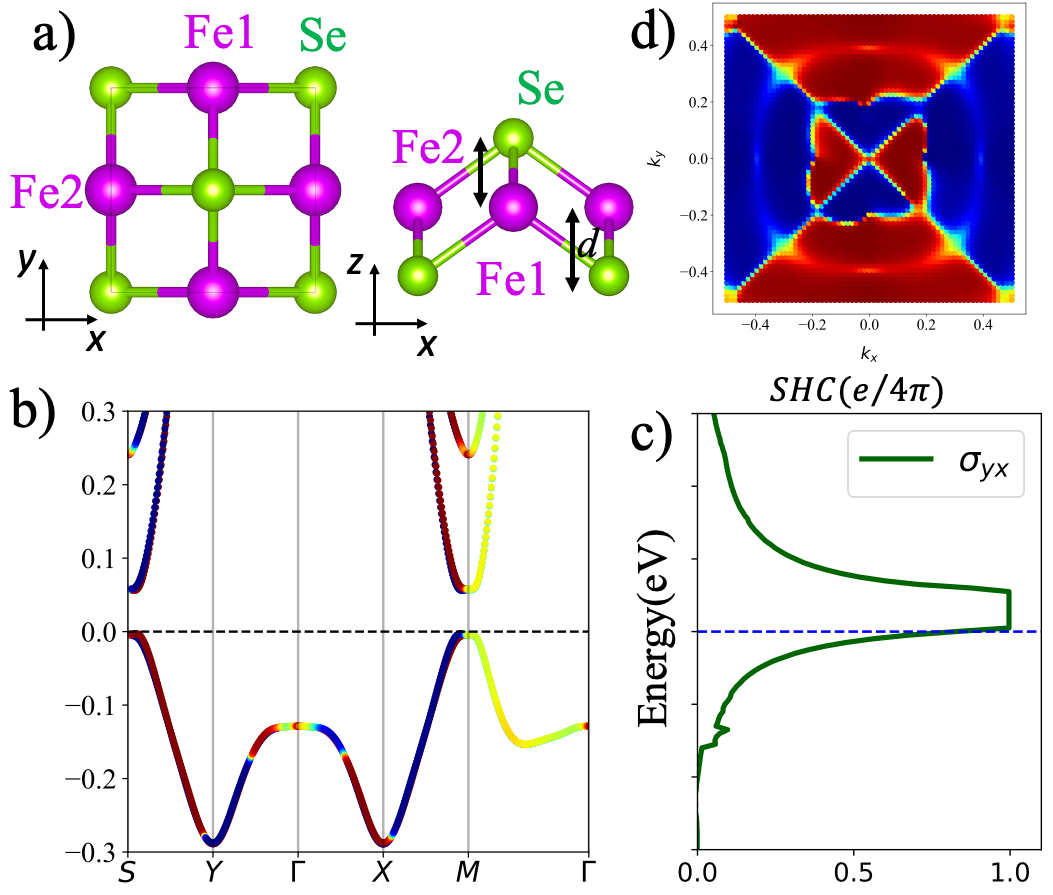}
	\caption{Electronic and topological properties of the two-dimensional altermagnetic phase in FeSe. (a) Ball-and-stick structural model, where Fe$_1$ and Fe$_2$ atoms have opposite magnetic moments. A slight vertical displacement of the bottom Se layer breaks $P\mathbb{T}$ symmetry. (b) Spin-resolved band structure showing momentum-dependent spin splitting with opposite signs along the $\Gamma$–X and $\Gamma$–Y directions. (c) Spin Hall conductivity $\sigma^z_{xy}$ as a function of Fermi energy, which presents a quantized value within the band gap. (d) Anisotropic $d$-wave spin-texture on the valence band. (e) Spin-projected surface states of the FeSe, revealing topological edge modes in the FeSe altermagnet phase.}
	\label{FeSe_material}
\end{figure}

Fig. \ref{FeSe_material} presents the material realization of a 2D-dimensional topological altermagnetic phase in the FeSe monolayer, a material recently studied and known for its structural simplicity \cite{fese-review,fese-dft,fese-super,fese-quantum}.  
The FeSe monolayer has a square Bravais lattice with two Fe atoms, and two Se atoms positioned above and below the Fe plane in a layered configuration.
Fig.~\ref{FeSe_material} a) illustrates the schematic model of the monolayer FeSe structure, where Fe$_1$ and Fe$_2$ atoms occupy distinct sublattices with antiparallel magnetic moments. 
The collinear antiferromagnetic configuration of monolayer FeSe yields zero net magnetization and preserves the combined inversion and time-reversal symmetries. 
This $P\mathbb{T}$ symmetry ensures a Kramers degeneracy at every $\mathbf{k}$-point in the Brillouin zone.

However, FeSe monolayer may realize an altermagnetic phase through various symmetry-breaking mechanisms~\cite{mazin2023fese}. 
In particular, when grown on a substrate as SrTiO$_3$ \cite{feseAFM_srtio3,fese_srtio3_nature}, vertical displacements of the Se atoms break inversion symmetry, thereby lifting the $P\mathbb{T}$ degeneracy. 
This symmetry breaking results in momentum-dependent spin splitting as shown in Fig.~\ref{FeSe_material} b).
The band spin-splitting changes sign between the $\Gamma$–X and $\Gamma$–Y directions, this anisotropic spin-splitting is the signal of a $d$-wave altermagnetic texture. 
This spin anisotropy is consistent with the $\mathcal{C}_{4z}\mathbb{T}$ symmetry explored in the Hamiltonian model of Eqn. \eqref{C4THamiltonian}.

Fig. \ref{FeSe_material} c-d) confirm the altermagnetic topological nature of the FeSe material. In Fig. \ref{FeSe_material} c), the spin Hall conductivity $\sigma^z_{xy}$ exhibits a quantized value at $e/4\pi$ within the bulk energy gap, indicating a nontrivial topological response associated with a spin Chern number of 1. 
This SCN=$1$ was independently calculated using the Chern number for the negative spin-valence eigenstates.
Fig. \ref{FeSe_material} d) illustrates the anisotropic $d$-wave spin texture of the valence band across the Brillouin zone, where the spin orientation varies in a $\mathcal{C}_{4z}\mathbb{T}$ symmetry shape.
Note the presence of an internal ring band inversion in the spin-resolved band structure, which is consistent with the predictions of the $\mathcal{C}_{4z}\mathbb{T}$ Hamiltonian model introduced previously in Eqn. \eqref{circles change of spin}.  
These results establish monolayer FeSe as a prototypical topological altermagnet, where $d$-wave spin splitting and quantized spin transport converge in a realistic 2D-dimensional material. 


\section{Conclusions}

The combination of rotational symmetries with time-reversal operations provides a framework for engineering topological phases in magnetic materials. 
By designing low-energy Hamiltonians with preserved $\mathcal{C}_{4z}\mathbb{T}$ and $\mathcal{C}_{3z}\mathbb{T}$ symmetries, 
we have demonstrated that nontrivial topological phases in 2D and 3D can be protected by these symmetries and characterized by spin Chern numbers, 
as predicted through K-theory analysis. 
Our findings show that altermagnetic systems, which exhibit non-relativistic spin splitting in the absence of net magnetization, can support topological phenomena, including symmetry-protected surface, hinge and corner states.
These boundary states can be manipulated with the local magnetic moments, represented by the $M_0$ term in the Hamiltonian. 
We have also proposed the FeSe monolayer as a realistic candidate for the experimental realization of 2D topological altermagnetic phases. 
This material can be classified as a $d$-wave altermagnetic materials with quantized spin Hall conductivity, where the topological  phase is protected by $\mathcal{C}_{4z}\mathbb{T}$.
Understanding topological altermagnets opens new pathways for spintronic technologies by connecting topological quantum effects with magnetic symmetries.

\vfill 

\section*{Acknowledgments}
RGH gratefully acknowledges the computing time granted on the supercomputer Mogon at Johannes Gutenberg University Mainz (hpc.uni-mainz.de). 
Additionally, the support from the Universidad Nacional de Colombia (QUIPU code 202010042199) and from MinCiencias through Convocatoria 937 for Fundamental Research is also deeply appreciated.
BU acknowledges the continuous support of the Max Planck Institute for Mathematics in Bonn, Germany,
and of the International Center for Theoretical Physics in Trieste, Italy, through its Associates Program.
RGH and BU thank the continuous support of the Alexander Von Humboldt Foundation, Germany.

\bibliographystyle{naturemag}
\bibliography{topological}

\end{document}